\documentclass[prb,unsortedaddress,twocolumn,superscriptaddress,9pt]{revtex4-1}

\pdfoutput=1
\usepackage{graphicx}
\DeclareGraphicsExtensions{ps,eps,ps.gz,frh.eps,ml.eps}
\usepackage{epstopdf}
\begin{document}


\newcommand{\Atilde}{\widetilde{A}}
\newcommand{\Etilde}{\widetilde{E}}
\newcommand{\w}{\omega}
\newcommand{\wopt}{\omega_{\rm opt}}
\newcommand{\wthz}{\omega_{\rm THz}}
\newcommand{\W}{\Omega}
\newcommand{\boldMatrix}[1]{ \underline{\underline{\mathbf{#1}}} }
\newcommand{\boldVect}[1]{ \underline{\mathbf{#1}} }
\newcommand{\chitwo}{\chi^{(2)}}
\newcommand{\chithree}{\chi^{(3)}}

\received{}
\accepted{}

\title{Revealing Carrier-Envelope Phase through Frequency Mixing and Interference in Frequency Resolved Optical Gating}

\author{E.W. Snedden}
	\email[]{edward.snedden@stfc.ac.uk}
\affiliation{Accelerator Science and Technology Center, STFC Daresbury National Laboratory,
Warrington, WA4 4AD, United Kingdom}
\author{D.A. Walsh}
\affiliation{Accelerator Science and Technology Center, STFC Daresbury National Laboratory,
Warrington, WA4 4AD, United Kingdom}
\author{S.P. Jamison}
\affiliation{Accelerator Science and Technology Center, STFC Daresbury National Laboratory,
Warrington, WA4 4AD, United Kingdom}
\affiliation{Photon Science Institute, The University of Manchester, Manchester M13 9PL, United Kingdom}

\date{\today}
\maketitle

\section{Abstract}

\noindent We demonstrate that full temporal characterisation of few-cycle electromagnetic pulses, including retrieval of the carrier envelope phase (CEP), can be directly obtained from Frequency Resolved Optical Gating (FROG) techniques in which the interference between non-linear frequency mixing processes is resolved.  We derive a framework for this scheme, defined Real Domain-FROG (ReD-FROG), as applied to the cases of interference between sum and difference frequency components and between fundamental and sum/difference frequency components.  A successful numerical demonstration of ReD-FROG as applied to the case of a self-referenced measurement is provided.  A proof-of-principle experiment is performed in which the CEP of a single-cycle THz pulse is accurately obtained and demonstrates the possibility for THz detection beyond the bandwidth limitations of electro-optic sampling.

\section{Introduction}

\noindent Few-cycle electromagnetic pulses offer a means of generating and controlling physical processes on sub-femtosecond timescales relevant in atomic and condensed matter physics.  The characterisation of such short pulses remains a crucial aspect of the experimental process and has a rich history~\cite{Rulliere2003,Trebino2009,Walmsley2009}.  
Of the many techniques available today, Frequency Resolved Optical Gating (FROG) has found widespread application due to its robustness and relatively simple experimental implementation.  An expansive array of related FROG techniques have been developed in the last two decades~\cite{Kane1993,Trebino1997,Trebino2002,OShea2001,Fittinghoff1996}; each offer specific advantages in varying experimental conditions, yet share a common mathematical and physical basis:  the measurement of an intensity spectrogram from a non-linear combination of pulses.  This is widely held to make determination of the carrier-envelope phase (CEP, $\phi^{CE}$) beyond FROG techniques
\cite{Trebino2009,Colberg2014,Nomura2013,Nomura2014,Cheng1999}. 

For electromagnetic pulses with many cycles of the carrier frequency the CEP approximates a simple time shift of the carrier oscillation and is often of limited physical interest; for few-cycle pulses however, CEP qualitatively alters the temporal profile of the electric field and subsequently the interaction physics of the pulse.  
In current ultrafast metrology full temporal characterisation typically requires a separate and often complex measurement of CEP in addition to information such as that provided by FROG.  Stereo-ATI has been used to measure CEP by measuring the direction of photoelectrons emitted following ionisation in a time-of-flight spectrometer.~\cite{Paulus2003,Paulus2005,Wittmann2009,Colberg2014}  CEP measurement has also been demonstrated in so-called attosecond streaking,~\cite{Hentschel2001,Kienberger2004} which utilises the spectral analysis of photoelectrons excited by sub-femtosecond high harmonic pulses.  In this latter case the principles of FROG have been applied to the emitted photoelectrons (as opposed to the input electromagnetic pulse as per usual) in the FROG-CRAB algorithm.~\cite{Mairesse2005}

The CEP appears as the zero-order (constant) term in a Taylor expansion of phase. If $\phi^{CE}$ is determined at one frequency it is therefore known uniquely. Recently, Nomura~\cite{Nomura2013} and Shira~\cite{Nomura2014} have utilised this, determining $\phi^{CE}$ at low frequencies through THz electro-optic sampling (EOS) techniques, obtaining complete pulse information when combined with traditional FROG techniques. 

Here we show that self and cross-referenced FROG techniques can be extended to be directly capable of full characterisation of the electric field temporal profile of few-cycle electromagnetic pulses; similarly, we show and experimentally demonstrate the complete retrieval of electric field can be obtained through cross-reference with higher frequency pulses, even when the higher-frequency probe is of similar duration and unknown carrier-envelope phase. There is no requirement for the to-be-determined pulse to contain low-frequency content accessible to $\delta$-function like sampling, with retrieval possible directly from frequency-mixing spectrograms.

\section{ReD-FROG: theory and retrieval algorithm}

\noindent FROG of electromagnetic pulse characterisation consists of a spectrally resolved auto- or cross-correlation of pulses mediated by a non-linear mixing process.  Herein the detailed discussion assumes second-order mixing through $\chi^{(2)}$, although the approach and results can be readily modified to higher order mixing. For observable, real, electric fields in the time domain, $E_{1}(t)$, $E_{2}(t)$ the complex spectra $\tilde{E}_{1}(\w)$, $\tilde{E}_{2}(\w)$ are defined from the Fourier transform, introducing the mathematical construct of negative frequencies.  For the $\chi^{(2)}$ interaction between pulses the spectrogram, the spectrum of $\chitwo$ generated field as a function of relative delay $\tau$ between input pulses, can be given as\cite{}:

\begin{eqnarray}
I(\w;\tau) && = \left| \tilde{R}(\w)\int_{-\infty}^{\infty}
{\rm d}\W \tilde{E}_1(\w-\W)\tilde{E}_2(\W)\exp(i\W\tau) \right|^2   ,
\label{spectrogram1:Eqn}
\end{eqnarray}
in which we take the common approximation that the linear and non-linear material response functions can be collected outside the integral in the response function $R(\w)$.~\cite{Baltuska1998,Cheng1999} Techniques such as surface harmonic generation \cite{Tsang1996, Anderson2008} or phase-match angle-dithering~\cite{Gu2002} are capable of very high bandwidth response, as would be necessary for few cycle pulses.  As the specific form of $R(\omega)$ does not affect either analysis or conclusion it is omitted from the following discussion.  

The carrier envelope phase $\phi^{CE}$ is the phase contribution that is widely accepted to be unmeasurable through FROG phase retrieval; to highlight conditions for observation and retrieval, $\phi^{CE}$ is explicitly separated out from the total spectral phase. All electric fields are explicitly constrained to be real in the time domain, which gives 
\begin{equation}
\tilde{E}_{\rm total}(\w) = \left\{ 
\begin{array}{ll}
\tilde{E}(\w)\exp(i\phi^{CE})   & ; \mbox{ }\w>0 \\
\tilde{E}^*(|\w|)\exp(-i\phi^{CE})   & ; \mbox{ }\w<0 \\
\end{array}\right.
\label{conjugate:Eqn}
\end{equation}
The imposition of the physical constraint for all fields to be purely real in the time domain and, in particular, the Heaviside functional form of $\phi^{CE}(\w)$, is what gives rise to the observability of the carrier-envelope phase.  Introducing the functional form of equation \ref{conjugate:Eqn} for the electric fields, the spectrogram can be expressed as
\begin{eqnarray}
I(\w;\tau) &  = &\left| SFG(\w;\tau) \right|^2 + \left| DFG_{+}(\w;\tau) \right|^2 +  \left| DFG_{-}(\w;\tau) \right|^2  \nonumber \\
\nonumber \\
&&  + 2\Re \left\{  SFG(\w;\tau)  DFG_{+}^*(\w;\tau) \,\,{\rm e}^{i2\phi^{CE}_2 }       \right\}  \nonumber \\ 
\nonumber \\
&&  + 2\Re \left\{  SFG(\w;\tau)  DFG_{-}^*(\w;\tau) \,\,{\rm e}^{i2\phi^{CE}_1}       \right\}  \nonumber \\
\nonumber  \\
&&  + 2\Re \left\{ DFG_{+}(\w;\tau) DFG_{-}^*(\w;\tau)   \,\,{\rm e}^{i2\phi^{CE}_1 - i2\phi^{CE}_2}       \right\} 
\label{spectrogram:Eqn}
\end{eqnarray}
where 
\begin{eqnarray}
SFG(\w;\tau) \equiv && \int_{0}^{\w} {\rm d}\W\, 
\tilde{E}_1(\w-\W)\tilde{E}_2(\W;\tau) , \nonumber \\
 %
DFG_{-}(\w;\tau) \equiv   && \int_{\w}^{\infty} {\rm d}\W\,
\tilde{E}_1^*(\W-\w)\tilde{E}_2(\W;\tau)
\nonumber \\
DFG_{+}(\w;\tau) \equiv  && \int_{0}^{+\infty} {\rm d}\W\,
\tilde{E}_1(\w+\W)\tilde{E}_2^*(\W;\tau) .
\label{sumdiffIntegrals:Eqn}
\end{eqnarray}
The contributions arising from sum-frequency ($SFG$) and difference frequency generation are represented by $SFG$ and ($DFG_{+}, DFG_{-}$) respectively.
When both sum and difference frequency contributions spectrally overlap non-zero the absolute carrier phase becomes observable.

\begin{figure}[htb]
	\includegraphics[width=8cm]{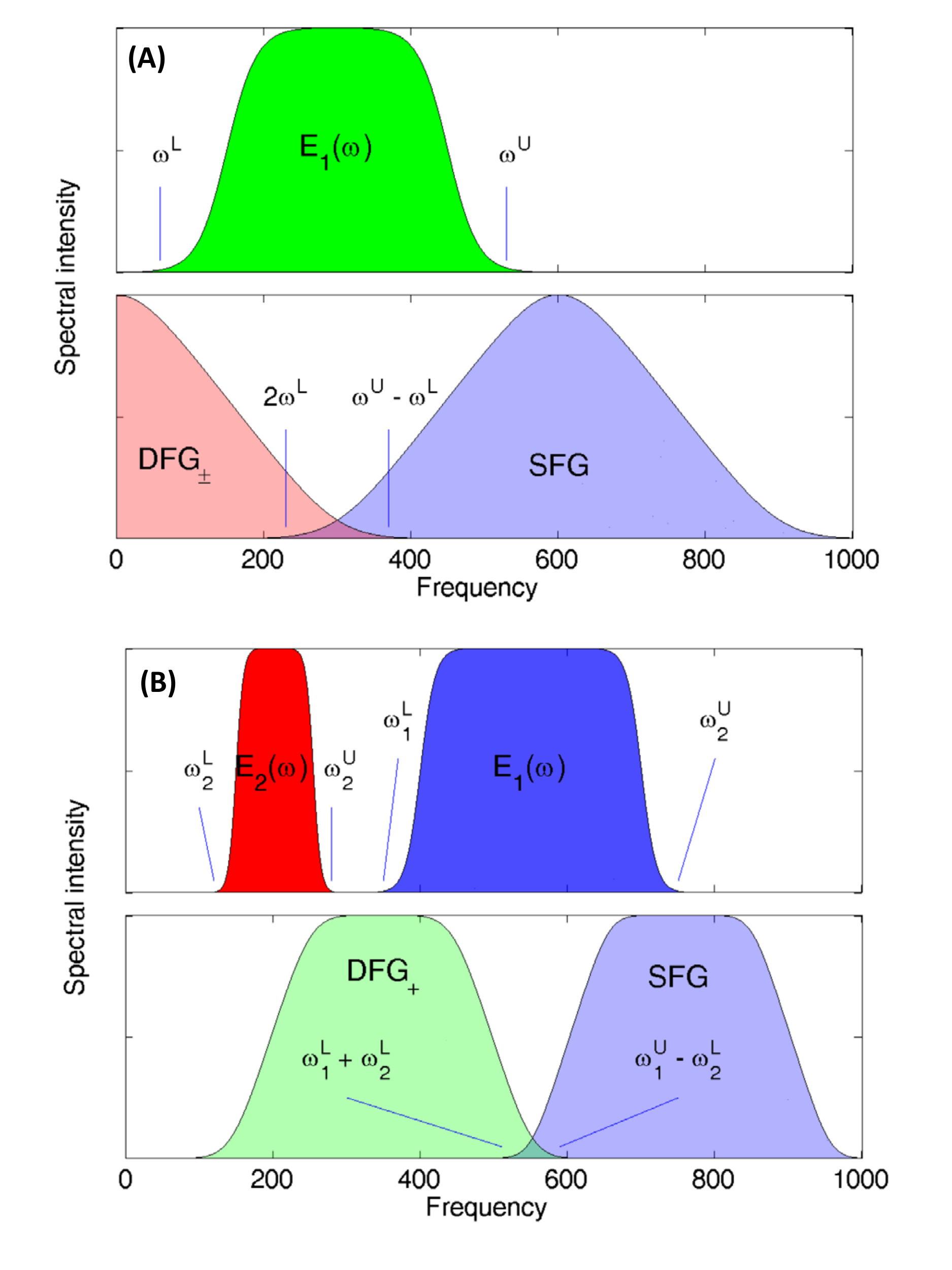}
	\caption{Schematic demonstrating bandwidth conditions for carrier envelope phase observation, for the two examples discussed in the text.  A) Self referenced mixing of a few cycle pulse. (top) spectrum for input pulse; (bottom) the sum and difference spectra. $DFG_+ = DFG_-$ for this configuration.
	(B) Cross-correlation of spectrally distinct pulses. (top) input spectra; (bottom) frequency mixed output, for which $DFG_- \equiv 0$.}
	\label{CEPoverlap:Fig}
\end{figure}

In this work we address two specific cases; that of a single broad-band pulse  cross-referenced with itself, and, as part of a proof-of-concept demonstration, the cross-correlation of two spectrally distinguishable pulses. The first case applies to the measurement of few-cycle optical pulses; the second case describes the unambiguous determination of the electric field of a pulse through sampling with another (unknown) optical field, which may be of similar or longer duration than the lower frequency pulse.  Representative field amplitude spectra of these two cases and the relative extent of the sum and difference frequency terms are shown schematically in Fig.~\ref{CEPoverlap:Fig}.  The principles discussed with these example can be readily extended to address other specific spectral considerations, such as spectrally overlapping pulses.

\begin{figure*}[htb]
			\includegraphics[width=14cm]{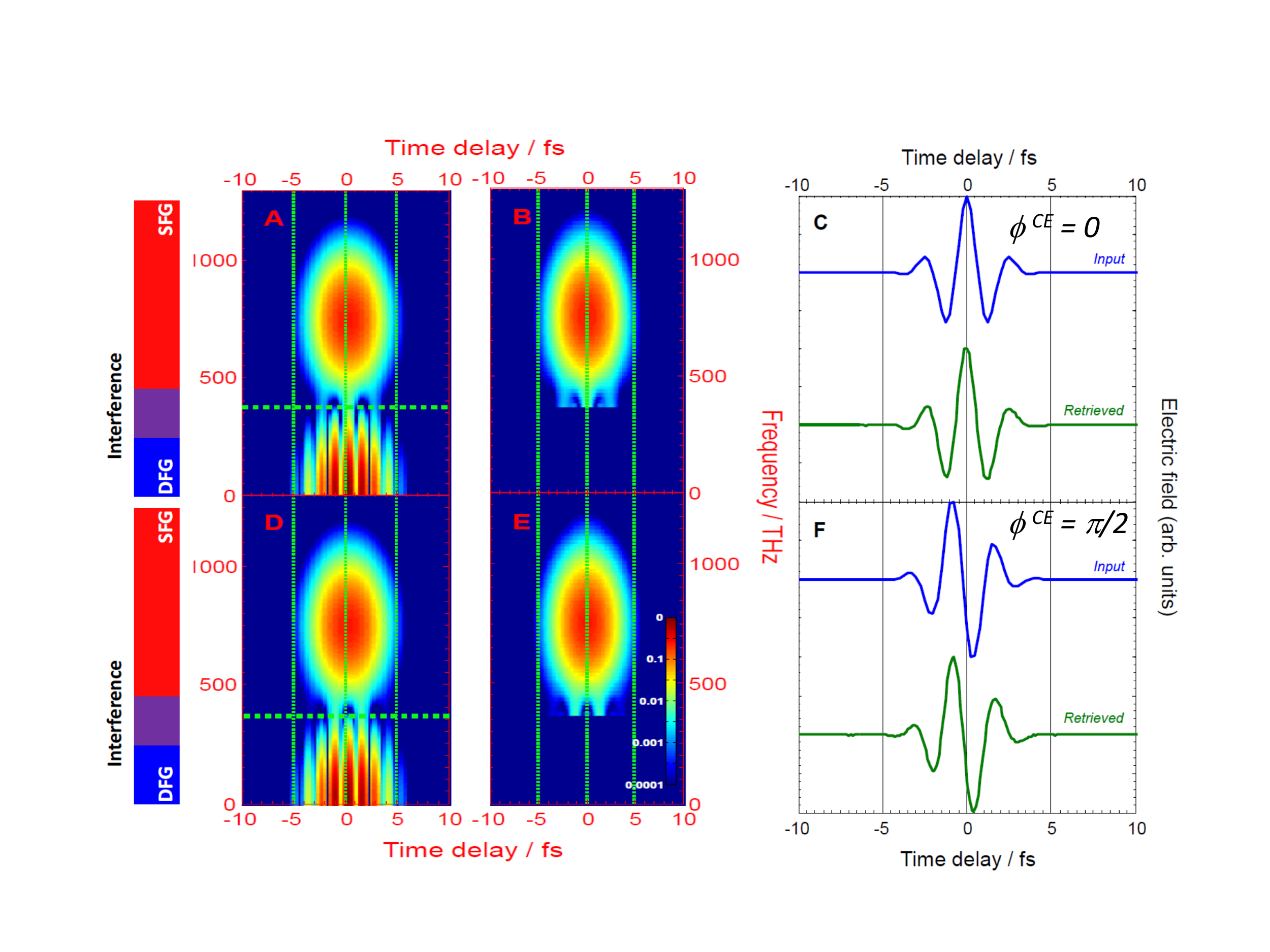}
	\caption{Numerical demonstration of the self-referencing retrieval of a transform-limited single-cycle optical pulse (110 THz bandwidth Gaussian with peak centred at 375 THz).  A) Simulated experimental spectrogram with $\phi^{CE}=0$. B) Retrieved spectrogram, in which the retrieval has been performed over a truncated spectral region (frequencies above 375 THz); C) comparison of input and retrieved electric field profile.  D-F) as above, with $\phi^{CE} = \pi/2 \mbox{ rad}$.
	All spectrograms are shown with logarithmic intensity scale.}
	\label{SRspec:Fig}
\end{figure*}

To demonstrate the complete electric field retrieval for the self-referenced measurement of a single pulse, test pulses with complex spectra differing only in carrier-envelope phase have been constructed, as shown in Fig.~\ref{SRspec:Fig}C, F. Simulated 'experimental' spectrograms were constructed directly as the product of the time domain fields. Modified FROG retrieval algorithms, accounting for the constraint of a strictly real time domain field,  are then applied to the 'experimental' spectrograms. Retrieval has been examined in presence of both measurement noise and spectral truncation.

Conventional FROG algorithms follow a two stage iterative procedure, the first being the comparison of the field amplitudes of the product spectrogram and the data. The second 'numerical-constraint' draws on the physics of the relevant frequency mixing process to extract the individual time domain fields from the field-product spectrogram. 
This constraint is conventionally undertaken from a functional minimisation between intensity spectrogram and the expected spectrogram inferred from iterated test fields and the physical process giving rise to frequency mixing.

We have modified the PCGPA FROG retrieval algorithm~\cite{Kane1999} to include the physical constraint of a real time-domain field, which together with the frequency-mixing spectral overlap, we label as ReD-FROG (Real Domain-FROG) retrieval for distinction from conventional algorithms.
The use of the PCGPA in providing the necessary deconvolution step decreases both the processing time and complexity of ReD-FROG, but is limited to measurements obeying equation \ref{spectrogram1:Eqn}.  For more complex pulses, or instances for which dispersion of the mixing medium cannot be neglected, a more complex version of the deconvolution algorithm may be required.~\cite{DeLong1995} 
The spectral constraint of real time-domain fields can be achieved through extending the measured spectrogram by mirror reflection about zero-frequency and applying the so-called 'numerical constraint' minimisation procedure on the extended data set. Through effective inclusion of negative frequencies, and the pinning of the spectrogram to have mirror symmetry, the input field retrieved is constrained to be real in the time domain.  In this case the mirror symmetry constrains the retrieved fields to have the Hermitian property necessary for the time domain fields to be purely real.

The retrieval algorithm was applied on a spectral truncation of the measurement data; such a spectral limited data set is to be expected in any practicable implementation.  To fully investigate the effect of truncation, retrieval was repeated for a range of cut-off frequencies between 325 and 375 THz.  
    
As shown in Fig.~\ref{SRspec:Fig}C, F, the correct and full (including CEP) electric field temporal profile is recovered from the modified retrieval algorithm. All ReD-FROG retrievals were based on a 512x512 grid. For Figs.~\ref{SRspec:Fig}C and \ref{SRspec:Fig}F, FROG errors of $5\times 10^{-3}$ and $6\times 10^{-3}$ were respectively obtained. 
In the given case the to-be-determined pulse was centred at 375\,THz, and only the spectrogram data for frequencies $>375$\,THz was included in the retrieval; similar agreement in the retrieved profile was obtained across the full range of cut-off frequencies investigated.  For an unambiguous electric field determination it is sufficient to include the sum-frequency region of the spectrogram, in the presence of some overlapping difference-frequency mixing spectral content.
Stable convergence was recorded with and without use of a spectral constraint corresponding to a measurement of spectral intensity~\cite{Trebino2002} applied over the truncated frequency range, although the rate of convergence was improved with its implementation.     

Up to this point in our analysis in  we have taken the conventional form of the FROG spectrogram (Equation~\ref{spectrogram1:Eqn}), in which the only the non-linear generated fields are included. 
In this arrangement the bandwidth ($\Delta \equiv \w^U-\w^L$) necessary for carrier envelope observability is $\Delta\geq2\w^L$. 
By inclusion of the input field in the measurement process, additional $\phi^{CE}$ dependent interference terms arise which allow the bandwidth requirements to be relaxed.  As a simplification, we assume that only one of the input fields spectrally overlaps with the mixing fields; under these conditions the spectrogram can be expressed as:
\begin{eqnarray}
I(\w;\tau) && = \left| \tilde{E}_1(\w) + \int_{-\infty}^{\infty}
{\rm d}\W \tilde{E}_1(\w-\W)\tilde{E}_2(\W)\exp(i\W\tau) \right|^2 .
\label{spectrogramf2f1:Eqn}
\end{eqnarray} 
Explicit introduction of the CEP and expansion of the integral (as performed for equation \ref{spectrogram:Eqn}) yields
\begin{eqnarray}
I(\w;\tau) &  = &\left| \tilde{E}_1(\w) \right|^2 + \left| SFG(\w;\tau) \right|^2 + \left| DFG_{+}(\w;\tau) \right|^2 +  \left| DFG_{-}(\w;\tau) \right|^2  \nonumber \\
\nonumber \\
&&  + 2\Re \left\{  \tilde{E}_1(\w)  SFG^*(\w;\tau) \,\,{\rm e}^{-i\phi^{CE}_2 }       \right\}  \nonumber \\ 
\nonumber \\
&&  + 2\Re \left\{  \tilde{E}_1(\w)  DFG_{+}^*(\w;\tau) \,\,{\rm e}^{i\phi^{CE}_2 }       \right\}  \nonumber \\ 
\nonumber \\
&&  + 2\Re \left\{  \tilde{E}_1(\w)  DFG_{-}^*(\w;\tau) \,\,{\rm e}^{i2\phi^{CE}_1 - i\phi^{CE}_2}       \right\}  \nonumber \\ \nonumber \\
&&  + 2\Re \left\{  SFG(\w;\tau)  DFG_{+}^*(\w;\tau) \,\,{\rm e}^{i2\phi^{CE}_2 }       \right\}  \nonumber \\ 
\nonumber \\
&&  + 2\Re \left\{  SFG(\w;\tau)  DFG_{-}^*(\w;\tau) \,\,{\rm e}^{i2\phi^{CE}_1}       \right\}  \nonumber \\
\nonumber  \\
&&  + 2\Re \left\{ DFG_{+}(\w;\tau) DFG_{-}^*(\w;\tau)   \,\,{\rm e}^{i2\phi^{CE}_1 - i2\phi^{CE}_2}       \right\} \ 
\label{spectrogramf2f:Eqn}
\end{eqnarray}

\begin{figure}[htb!]
	\includegraphics[width=6cm]{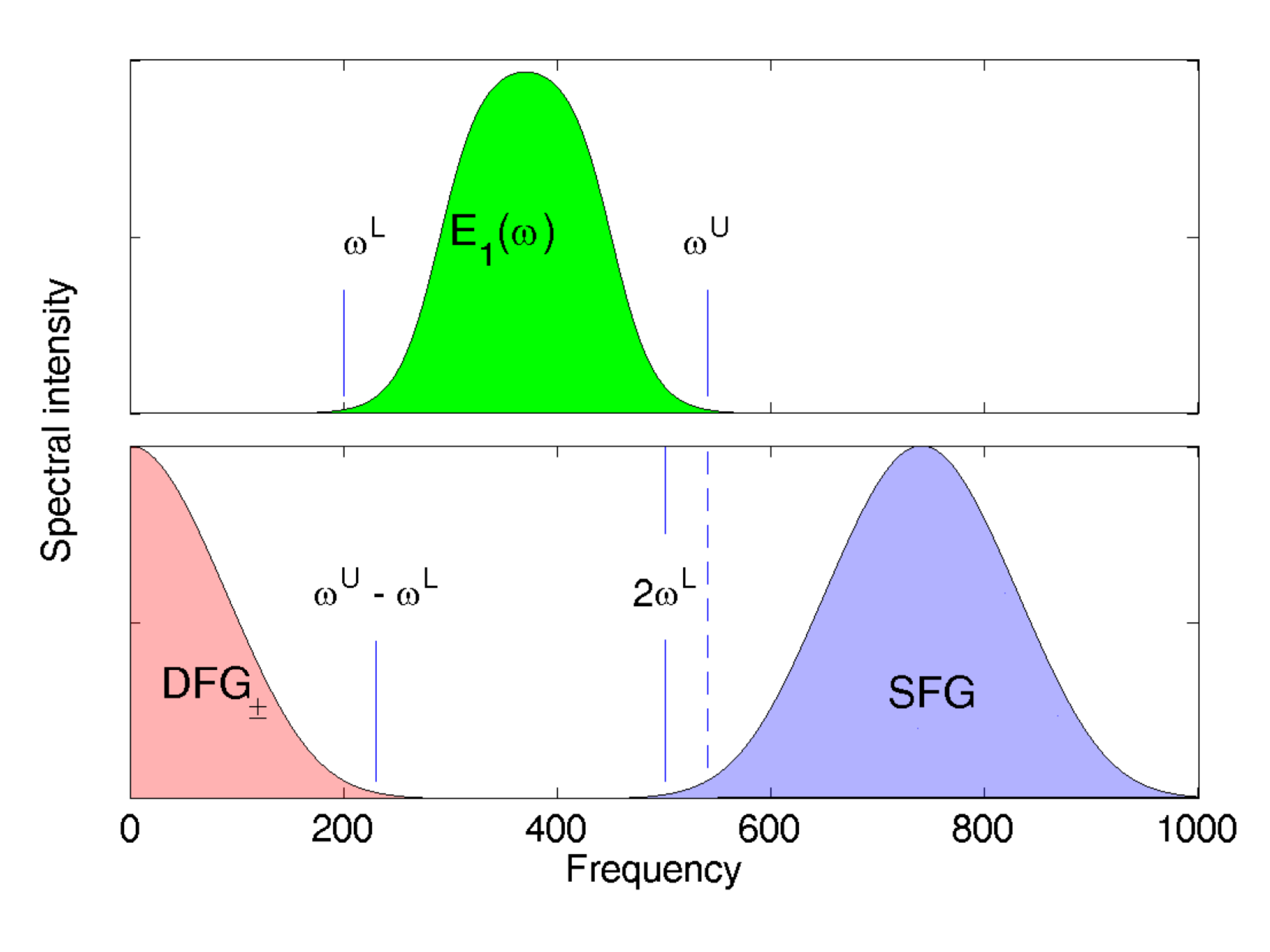}
	\includegraphics[width=7cm]{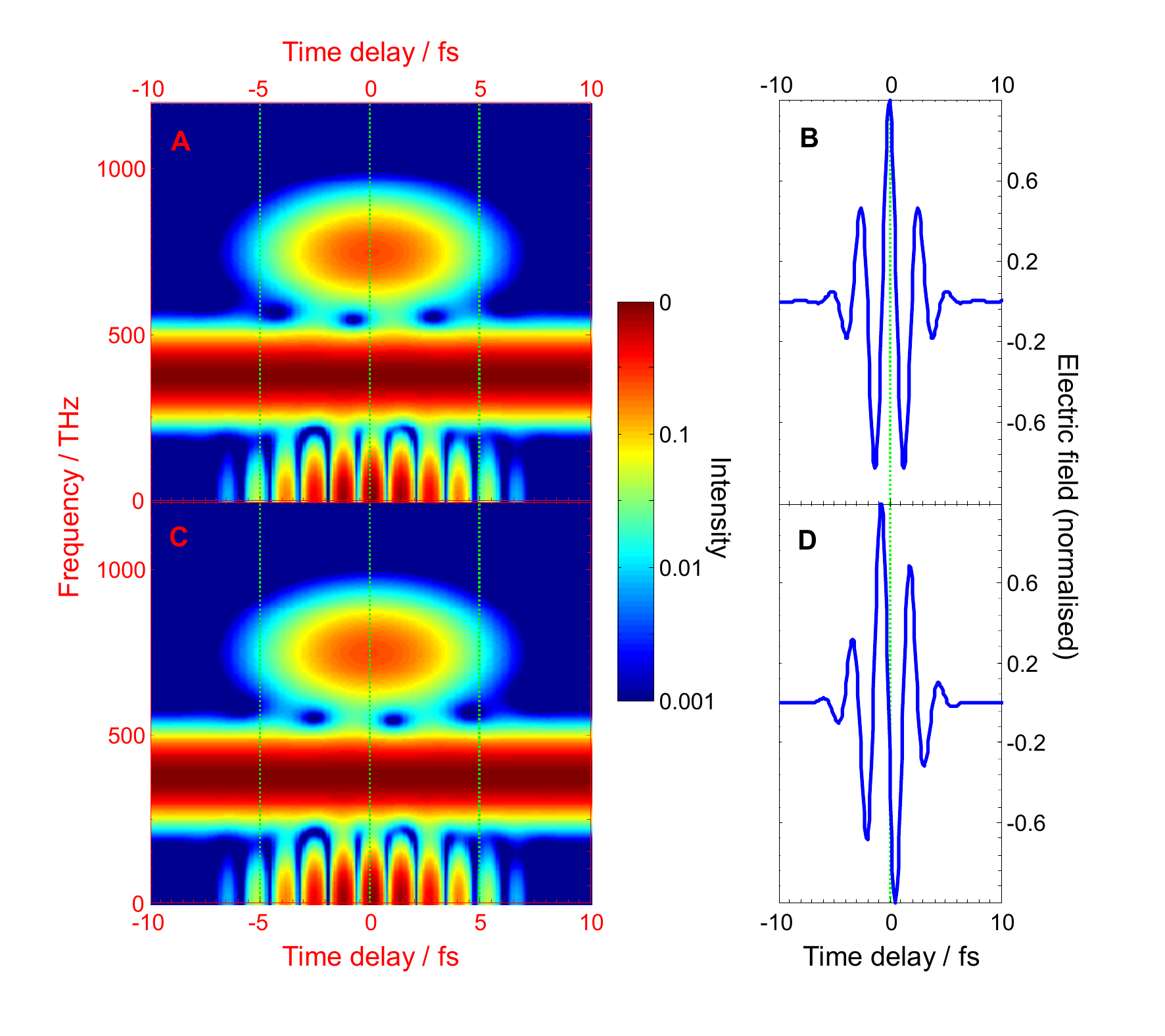}
	\caption{Schematic of bandwidth requirements for carrier phase observation through interference between fundamental and sum (or difference) frequency fields. 
	(B,C) Spectrograms for a self-referencing measurement of a Gaussian transform-limited femtosecond pulse (375 THz carrier frequency, bandwidth 75 THz), $\phi^{CE} = 0$ and $\phi^{CE} = \pi/2 \mbox{ rad}$ for (B) and (C) respectively.}
	\label{SRspec2:Fig}
\end{figure}

The additional interference terms in the above expression allow for carrier envelope phase observation even when there is no spectral overlap between sum and difference frequency mixing.
The first two interference terms in equation \ref{spectrogramf2f:Eqn} yield measurement of the phase  $\phi^{CE}_2$ through interference between either the fundamental field $\tilde{E}_1$ and the sum-frequency field, or between the fundamental and $\tilde{E}_1$ and difference frequency fields. The relative bandwidth requirements are shown schematically in Fig.~\ref{SRspec2:Fig}A; the carrier envelope phase is observable for a bandwidth satisfying $\Delta\geq\w^L$.  
This criterion is the same as that utilised in f-2f beating schemes for the CEP stabilisation of repetitive pulse trains;~\cite{Telle1999,Jones2000,Holzworth2000,Paulus2005} such stabilisation schemes however cannot provide a practical measurement of CEP, requiring extremely accurate characterisation of the dispersion associated with the experimental apparatus.~\cite{Paulus2005,Walmsley2009}    Equation \ref{spectrogramf2f:Eqn} demonstrates resolving this process as part of a FROG measurement provides the means to measure absolute phase without the need of a frequency comb.   

Figure \ref{SRspec2:Fig}B presents an example of the spectrogram arising from a self-referencing measurement of a transform-limited Gaussian pulse (75 THz bandwidth) in which overlap with the input field occurs.  The overlap between sum and difference frequency generated fields is absent in this case and the interference features arise from the first three interference terms in equation \ref{spectrogramf2f:Eqn}.
 
\section{Experiment: characterisation of single-cycle THz radiation}

\noindent We now consider the ReD-FROG process applied to the case of spectrally distinct pulses, as in Figure \ref{CEPoverlap:Fig}B. For this configuration $DFG_{-}(\w;\tau)\equiv 0$, and therefore $\phi^{CE}_1$ and  $\phi^{CE}_1-\phi^{CE}_2$ are not observable. Conversely the carrier envelope phase $\phi^{CE}_2$ observable is independent of the (unknown) CEP phase of the upper band pulse.  To observe $\phi^{CE}_2$ we require $\Delta_1 \equiv \w^U_1-\w^L_1 > \w^L_2$. 

In this case a benchmarked proof-of-concept experiment is feasible; a quasi-single cycle THz pulse can serve as the lower frequency pulse to be determined, while an optical pulse of similar bandwidth acts as the higher frequency probe. For the benchmarking, a significantly broader bandwidth $\delta$-function like optical pulse can be employed for electro-optic sampling (EOS),~\cite{Gallot1999,Jamison2006} obtaining an independent measure of the true electric field profile.  

\begin{figure}[htb!]
	\includegraphics[width=9cm]{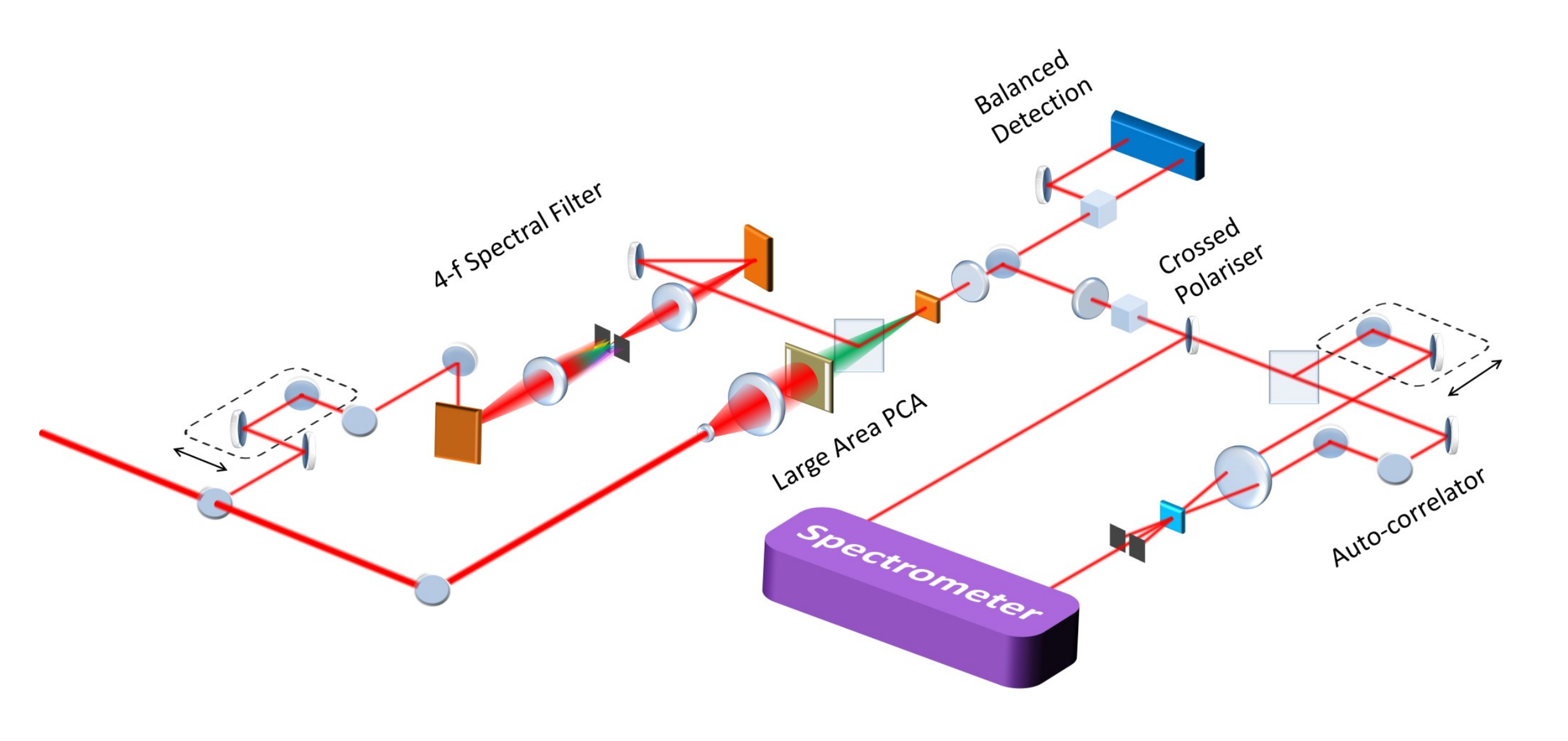}	
	\includegraphics[width=9cm]{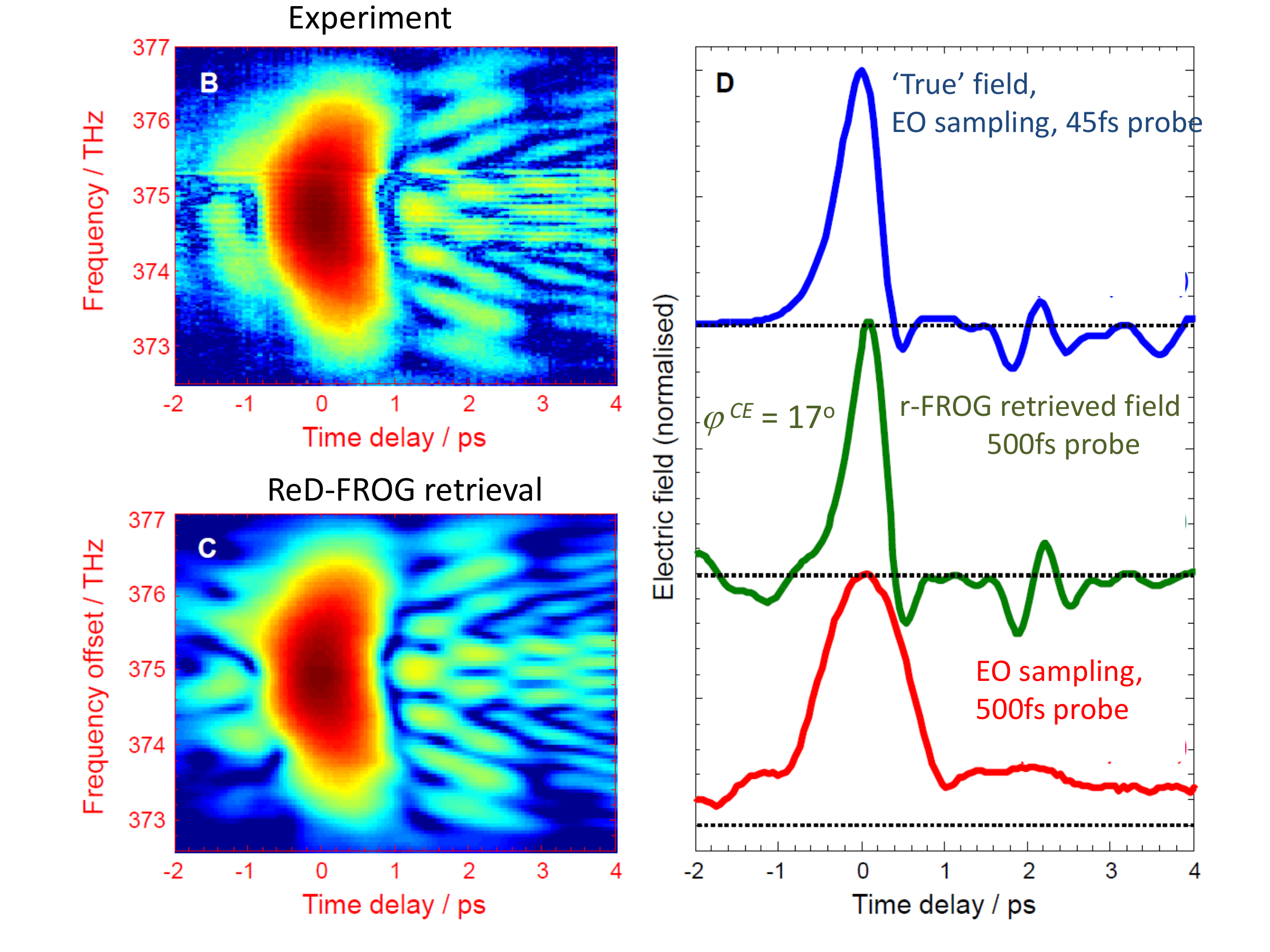}
	\caption{A) Schematic of experimental arrangement for carrier phase retrieval and benchmarking.
	 B) Measured spectrogram with a 1\,THz bandwidth, 500\,fs duration, optical probe; C) ReD-FROG retrieved spectrogram  D) comparison of true electric field as determined by electro-optic sampling with a 45\,fs probe, and that from the ReD-FROG retrieval.  The electro-optic sampling measurement for the spectrally narrowed 500\,fs probe is also shown for comparison. 	Spectrograms are shown with logarithmic intensity scale.}
	\label{Experiment:Fig}
\end{figure}

Such an experimental demonstration has been performed, with the experimental arrangement shown in Fig.~\ref{Experiment:Fig}. An amplified femtosecond Ti:Sapphire laser system (Coherent Micra/Legend, amplifier output: 45\,fs, 1\,mJ at 800\,nm) was used to drive a large-area semi-insulating GaAs photo-conductive antenna, producing quasi-single cycle sub-ps THz pulses.  The pulses were detected within the near-field region of the antenna, allowing the electric field oscillation to have a unipolar appearance.
The optical probe pulse was selected prior to the PCA using a beam-splitter and directed through a zero-dispersion 4-f filter, allowing the pulse bandwidth to be selected as required.  The benchmarking measurement of the temporal profile of the THz pulse was performed using a standard balanced detection EOS arrangement using a 45\,fs optical probe.~\cite{Walsh2014}  For the spectrogram retrieval experiments the optical bandwidth was restricted to 1\,THz, less than the full bandwidth of the THz pulse, with corresponding increase in duration to 500\,fs.
Optical probe temporal envelopes were confirmed with a custom-built SHG-FROG setup with 100 $\mu$m thick BBO.

Both optical and THz beams were focused into a 
$\langle\bar{1}10\rangle$ orientated ZnTe crystal.  The frequency-mixing signal was isolated from residual input-probe light using a Glan-Thompson calcite polariser and spectrally analysed as a function of the relative delay between the THz and optical probe beams using an imaging spectrometer (Jobin Yvon, iHR550) and intensified CCD camera (PCO DiCamPRO).  A ReD-FROG algorithm as described earlier was applied to retrieve the field, using only the measured spectrogram as input.

The measured and retrieved spectrograms are shown in Figures \ref{Experiment:Fig}B,C, together with the ReD-FROG retrieved field time profile.  The FROG error in the retrieved spectrogram was 0.01.  
  Despite the complex form of the THz pulse, both main peak and oscillations over a period of several picoseconds (due to THz absorption by ambient water vapour) are recovered almost perfectly and the CEP of the single-cycle pulse is correctly identified.  
  For comparison the field profile inferred by direct electro-optic sampling with the 1\,THz bandwidth probe, without spectrally resolved phase retrieval, is also shown. 
  
  Whilst this experiment confirms the validity of CEP retrieval using ReD-FROG, the approach also finds immediate application in surpassing probe-duration limitations present in high time-resolution EOS.  Continuing the notation of section II, these limitations can be expressed in the frequency domain as $\Delta\gg\w^H$; for the same optical probe duration, we estimate that ReD-FROG ($\Delta\geq\w^L$) is capable of measuring THz pulse durations $\times$3-4 shorter than that accessible through EOS.   The only existing or proposed THz schemes with similar capabilities, such as spectral upconversion~\cite{Jamison2010} or BMX-FROG,~\cite{Helle2012} rely on multi-stage detection in which the output of the electro-optic effect is measured as part of a second, distinct non-linear process (such as an auto-correlation or FROG).  The use of consecutive non-linear stages for pulse analysis limits the available signal to noise ratio and thus the range of THz sources that can be measured; the application of the ReD-FROG method, being a single-stage direct analysis of the EO effect, overcomes such limitations and presents as a more robust and versatile measurement technique.

\section{Conclusions}

\noindent We have demonstrated that a spectrally-resolved measurement of the superposition of non-linear frequency mixing processes is capable of unambiguous electric field temporal characterisation, including the measurement of carrier envelope phase.  Numerical simulations have been used to demonstrate the capability for few-cycle optical pulses.  A proof-of-concept benchmarked experiment using single-cycle THz radiation was performed and confirmed the validity of the formalism.  This method opens up new avenues of possibility in the measurement of few-cycle electric fields and demonstrates that, contrary to previous expectations, FROG is directly capable of measuring absolute phase.  Potential applications for ReD-FROG in high time-resolution electro-optic sampling are also envisaged as a means to exceed the bandwidth limitations of standard electro-optic sampling techniques.




\clearpage


\end{document}